\documentstyle[sprocl,epsf]{article}

\begin{document}

\title{			A PHENOMENOLOGICAL OUTLOOK \\
		ON THREE-FLAVOR ATMOSPHERIC NEUTRINO 
			    OSCILLATIONS}

\author{	G.L.\ Fogli, E.\ Lisi, A.\ Marrone, and G.\ Scioscia}

\address{	Dipartimento di Fisica e INFN, Sezione di Bari\\
			Via Amendola 173, 70126 Bari, Italy} 

\maketitle
\abstracts{The recent observations of atmospheric $\nu$ events from the
Super-Kamiokande experiment are compatible with three-flavor neutrino
oscillations, occurring dominantly in the $\nu_\mu\leftrightarrow\nu_\tau$
channel and subdominantly in the $\nu_\mu\leftrightarrow\nu_e$ channel. We
present an updated analysis of the three-flavor mass-mixing parameters 
consistent with the present phenomenology, including the latest 45 kTy data
sample from Super-Kamiokande. A comparison with our previous results, based on
33 kTy data, shows that the oscillation evidence is strengthened, and that the
neutrino mass-mixing parameters are constrained in smaller ranges.}

\section{Introduction}

The recent atmospheric neutrino data from the Super-Kamiokande (SK) experiment
\cite{Su99} are in excellent agreement with the hypothesis of flavor
oscillations generated by nonzero neutrino mass and mixing \cite{Po67} in the
$\nu_\mu\leftrightarrow\nu_\tau$ channel \cite{SKEV}. Such hypothesis is
consistent with all the SK data, including sub-GeV  $e$-like and $\mu$-like
events (SG$e,\mu$) \cite{SKSG}, multi-GeV  $e$-like and $\mu$-like events
(MG$e,\mu$) \cite{SKMG}, and upward-going muon events (UP$\mu$) \cite{SKUP},
and is also corroborated by independent atmospheric neutrino results from the
MACRO \cite{MACR} and Soudan-2 \cite{SOUD} experiments, as well as by the
finalized upward-going muon data sample from the pioneering Kamiokande
experiment \cite{KAUP}. Oscillations in the $\nu_\mu\leftrightarrow\nu_\tau$
channel are also compatible with the negative results of the reactor experiment
CHOOZ in the $\nu_\mu\leftrightarrow\nu_e$ channel \cite{CHOO}.

However, it has been realized that {\em dominant\/}
$\nu_\mu\leftrightarrow\nu_\tau$ oscillations plus {\em subdominant\/}
$\nu_\mu\leftrightarrow\nu_e$  oscillations are also consistent with SK+CHOOZ
data, and lead to a much richer three-flavor oscillation phenomenology
\cite{Fo99}.  A detailed $3\nu$ analysis of the SK observations, including the
full 33 kTy data sample, can be found in Ref.~\cite{Fo99}. Here we report and
comment briefly the results of our updated analysis, based on the recent  45
kTy SK data  \cite{Me99,Ha99}. The theoretical framework is based on the
so-called one mass scale dominance \cite{Fo95}, which has been used also for 
three-flavor oscillation studies of pre-SK  atmospheric and reactor neutrino
data in Refs.~\cite{Fo95,Fo97,Fo98}.

\section{$3\nu$ analysis of SK phenomenology (45 kTy)}

In the hypothesis that the two lightest neutrinos $(\nu_1,\nu_2)$ are
effectively degenerate $(m^2_1\simeq m^2_2)$ (one mass scale dominance), it can
be shown \cite{Fo95,Fo99} that atmospheric neutrinos probe only $m^2\equiv
m^2_3-m^2_{1,2}$, together with mixing matrix elements $U_{\alpha i}$ related
to $\nu_3$, namely: \begin{equation} {\rm Parameter\ space}
\;\equiv\;(m^2,U^2_{e1},U^2_{e2},U^2_{e3})\ , \end{equation} where
$U^2_{e1}+U^2_{e2}+U^2_{e3}=1$ for unitarity. The unitarity constraint can be
conveniently embedded in a triangle plot \cite{Fo95,Fo97,Fo99}, whose corners
represent the flavor eigenstates, while the heights projected from any inner
point  represent  the $U^2_{\alpha3}$'s.

Within this framework, we analyze 30 data points, related to the zenith
distributions of sub-GeV events (SG $e$-like and $\mu$-like, 5+5 bins), 
multi-GeV events (MG$e$,$\mu$ 5+5 bins) and upward-going muons  (UP$\mu$, 10
bins), using the latest 45 kTy SK sample \cite{Me99,Ha99}. We also consider the
rate of events in the CHOOZ reactor experiment \cite{CHOO}  (one bin), which
constrains the $\nu_e$ disappearance channel. Constraints are obtained through
a $\chi^2$ statistic, as described in Ref.~\cite{Fo99}.

Figure~1 shows the regions favored at 90\% and 99\% C.L.\ in the triangle plot,
for representative values of $m^2$. The CHOOZ data, which exclude a large
horizontal strip in the triangle, appear to be crucial in constraining
three-flavor mixing. Pure $\nu_\mu\leftrightarrow\nu_e$ oscillations (right
side of the triangles) are excluded by SK and CHOOZ independently.  The center
of the lower side, corresponding to pure $\nu_\mu\leftrightarrow\nu_\tau$
oscillations with maximal mixing, is allowed in each triangle both by SK and
SK+CHOOZ data. However, deviations from maximal
$(\nu_\mu\leftrightarrow\nu_\tau)$ mixing, as well as subdominant mixing with
$\nu_e$, are also allowed to some extent. Such deviations from maximal  $2\nu$
mixing  are slightly more constrained than  in the previous analysis of the 33
kTy SK data \cite{Fo99}.

Figure~2 shows the constraints on the mass parameter $m^2$ for unconstrained
three-flavor mixing. The best fit value is reached at $m^2\sim3\times 10^{-3}$
eV$^2$, and is only slightly influenced by the inclusion of CHOOZ data.
However, the upper bound on $m^2$ is significantly improved by including CHOOZ.
As compared with the corresponding plot in Ref.~\cite{Fo99} (33 kTy), this
figure shows that the 45 kTy data are in better agreement with the oscillation
hypothesis (lower $\chi^2$). Moreover, the favored range of $m^2$ is restricted
by $\sim 10\%$ with respect to Ref.~\cite{Fo99}.

Figures~1 and 2 clearly show the tremendous impact of the SK experiment in
constraining the neutrino oscillation parameter space. Prior to SK, the data
could not significantly favor  $\nu_\mu\leftrightarrow\nu_\tau$ over
$\nu_\mu\leftrightarrow\nu_e$  oscillations, and could only  put weak bounds on
$m^2$ (see Refs.~\cite{Fo97,Fo98}).

\begin{figure}
\epsfysize=14.5truecm
\hspace*{0.7truecm}
\epsfbox{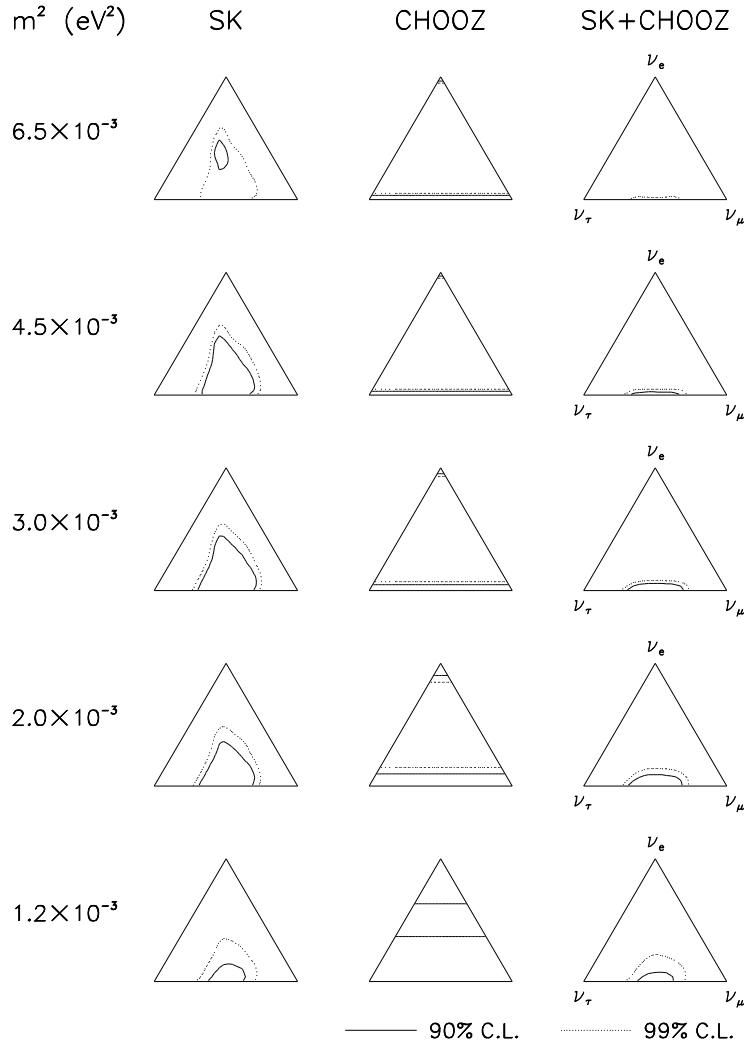}
\vskip-1.3cm
\caption{Three-flavor analysis in the triangle plot, for five representative
values of $m^2$. Left and middle column: separate analyses of Super-Kamiokande
(45 kTy) and CHOOZ data, respectively. Right column: combined SK+CHOOZ allowed
regions. Although the SK+CHOOZ solutions are close to pure
$\nu_\mu\leftrightarrow\nu_\tau$ oscillations, the allowed values of $U^2_{e3}$
are never negligible, especially in the lower range of $m^2$.}
\label{fig1}
\end{figure}
\newpage

\begin{figure}
\epsfysize=9.truecm
\phantom{.}
\vspace*{-.5truecm}
\hspace*{2.3truecm}
\epsfbox{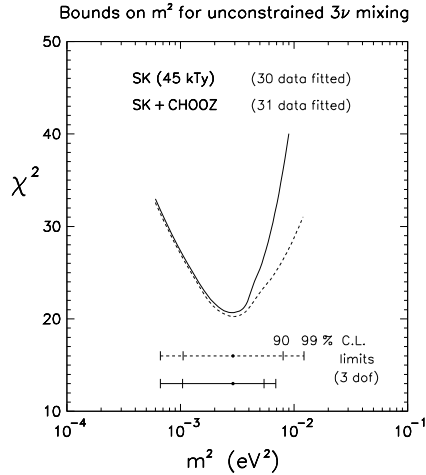}
\vskip-2.5cm
\caption{Values of $\chi^2$ as a function of $m^2$, for unconstrained
three-flavor mixing. Dashed curve: fit to SK data only (45 kTy). Solid curve:
fit to SK+CHOOZ. The minimum of $\chi^2$ is reached for  $m^2\simeq 2.8\times
10^{-3} {\rm\ eV}^2$.}
\label{fig2}
\end{figure}

Finally, Fig.~3 shows the best-fit zenith distributions of SG$e,\mu$,
MG$e,\mu$, and UP$\mu$ events, normalized to the no-oscillation rates in each
bin. There is excellent agreement between data and theory, especially for the
$\mu$ distributions.  The nonzero value of $U^2_{e3}$ at best fit leads to a
slight expected electron excess in the MG$e$ sample for $\cos\theta\to -1$. The
observed electron excess is, however, somewhat larger than expected, both for
SG$e$'s and for MG$e$'s. A significant reduction of the statistical error is
needed to probe possible MG$e$ distortions,  which would be  unmistakable
signals of subdominant $\nu_\mu\to\nu_e$ oscillations.

\section{Outlook}

The Super-Kamiokande data are compatible with three-flavor oscillations
dominated by $\nu_\mu\leftrightarrow\nu_\tau$ transitions. The amplitude of the
$\nu_\mu\leftrightarrow\nu_e$ channel is not necessarily zero, although being
strongly constrained  by both SK and  CHOOZ. A contribution from the
$\nu_\mu\leftrightarrow\nu_e$ channel might explain part of the electron excess
observed in SK, especially for multi-GeV $e$-like events. Higher statistics is
needed to test such interpretation. A definite  progress in confirming the
oscillation hypothesis, and in constraining the mass-mixing parameters, emerges
from a comparison of the 33 kTy and 45 kTy SK data analyses.\\[2mm]
{\bf Acknowledgments.} G.L.F.\ thanks the organizers of the workshop WIN'99
for kind hospitality.

\newpage

\begin{figure}
\epsfysize=12.truecm
\phantom{.}
\vspace*{-2.8truecm}
\hspace*{1.3truecm}
\epsfbox{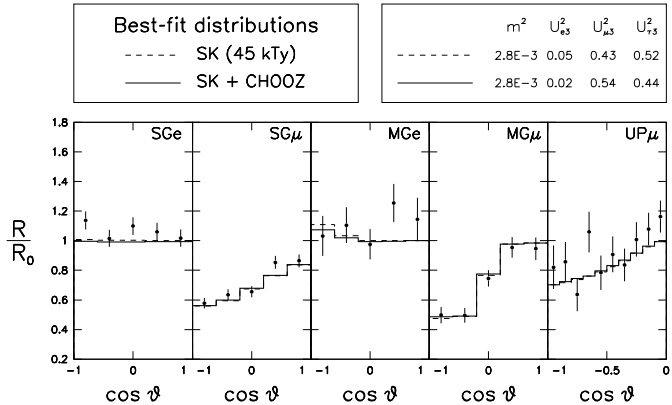}
\vskip-4.0cm
\caption{SK zenith distributions of leptons at best fit (dashed lines), also
including CHOOZ (solid lines), as compared with the 45 kTy experimental data 
(dots with error bars). The $3\nu$ mass-mixing values at best fit are indicated
in the upper right corner.}
\label{fig3}
\end{figure}


\end{document}